# Network Coding for Distributed Storage Systems

Alexandros G. Dimakis, P. Brighten Godfrey, Yunnan Wu,
Martin O. Wainwright and Kannan Ramchandran
Department of Electrical Engineering and Computer Science,
University of California, Berkeley, CA 94704.
Email: {adim, pbg, wainwrig, kannanr}@eecs.berkeley.edu and yunnanwu@microsoft.com

***Abstract*—Distributed storage systems provide reliable access to data through redundancy spread over individually unreliable nodes. Application scenarios include data centers, peer-to-peer storage systems, and storage in wireless networks. Storing data using an erasure code, in fragments spread across nodes, requires less redundancy than simple replication for the same level of reliability. However, since fragments must be periodically replaced as nodes fail, a key question is how to generate encoded fragments in a distributed way while transferring as little data as possible across the network.**

**For an erasure coded system, a common practice to repair from a node failure is for a new node to download subsets of data stored at a number of surviving nodes, reconstruct a lost coded block using the downloaded data, and store it at the new node. We show that this procedure is sub-optimal. We introduce the notion of regenerating codes, which allow a new node to download *functions* of the stored data from the surviving nodes. We show that regenerating codes can significantly reduce the repair bandwidth. Further, we show that there is a fundamental tradeoff between storage and repair bandwidth which we theoretically characterize using flow arguments on an appropriately constructed graph. By invoking constructive results in network coding, we introduce regenerating codes that can achieve any point in this optimal tradeoff.**

## I. Introduction

The purpose of distributed storage systems is to store data reliably over long periods of time using a distributed collection of storage nodes which may be individually unreliable. Applications involve storage in large data centers and peer-to-peer storage systems such as OceanStore [3], Total Recall [4], and DHash++ [5], that use nodes across the Internet for distributed file storage. In wireless sensor networks, obtaining reliable storage over unreliable motes might be desirable for robust data recovery [6], especially in catastrophic scenarios [7].

In all these scenarios, ensuring reliability requires the introduction of redundancy. The simplest form of redundancy is replication, which is adopted in many practical storage systems. As a generalization of replication, erasure coding offers better storage efficiency. For instance, we can divide a file of size $\mathcal{M}$ into $k$ pieces, each of size $\mathcal{M}/k$, encode them into $n$ coded pieces using an $(n, k)$ maximum distance separable (MDS) code, and store them at $n$ nodes. Then, the original file can be recovered from any set of $k$ coded pieces.

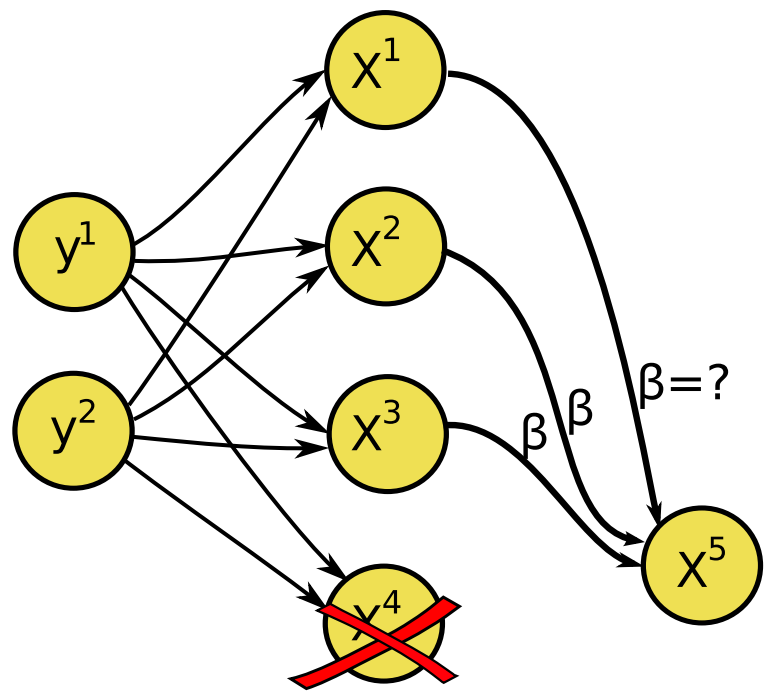

Fig. 1. The repair problem: Assume that a (4,2) MDS erasure code is used to generate 4 fragments (stored in nodes $x^1, \ldots x^4$) with the property that any 2 can be used to reconstruct the original data $y^1, y^2$. When node $x^4$ fails, and a newcomer $x^5$ needs to generate an erasure fragment from $x^1, \ldots x^3$, what is the minimum amount of information that needs to be communicated?

This performance is optimal in terms of the redundancy–reliability tradeoff because $k$ pieces, each of size $\mathcal{M}/k$, provide the minimum data for recovering the file, which is of size $\mathcal{M}$. Several designs [8], [4], [5] use erasure codes instead of replication. For certain cases, erasure coding can achieve orders of magnitude higher reliability for the same redundancy factor compared to replication; see, e.g., [9].

However, a complication arises: In distributed storage systems, redundancy must be continually refreshed as nodes fail or leave the system, which involves large data transfers across the network. This problem is best illustrated in the simple example of Fig. 1: a data object is divided in two fragments $y^1, y^2$ (say, each of size 1Mb) and these encoded into four fragments $x^1, \ldots x^4$ of same size, with the property that any two out of the four can be used to recover the original $y^1, y^2$. Now assume that storage node $x^4$ fails and a new node $x^5$, the newcomer, needs to communicate with existing nodes and create a new encoded packet, such that any two out of $x^1, x^2, x^3, x^5$ suffice to recover. Clearly, if the newcomer can download any two encoded fragments (say from $x^1, x^2$), reconstruction of the whole data object is possible and then a new encoded fragment can be generated (for example by making a new linear combination that is independent from the existing ones). This, however, requires the communication of 2Mb in the network to generate an erasure encoded fragment of size 1Mb at $x^5$. In general, if an object of size $\mathcal{M}$ is divided in $k$ initial fragments, the repair bandwidth with this strategy is $\mathcal{M}$ bits to generate a fragment of size $\mathcal{M}/k$. In contrast, if replication is used instead, a new replica may simply be copied from any other existing node, incurring no bandwidth overhead. It was commonly believed that this $k$-factor overhead in repair bandwidth is an unavoidable overhead that comes with the benefits of coding (see, for example, [10]). Indeed, all known coding constructions require access to the original data object to generate encoded fragments.

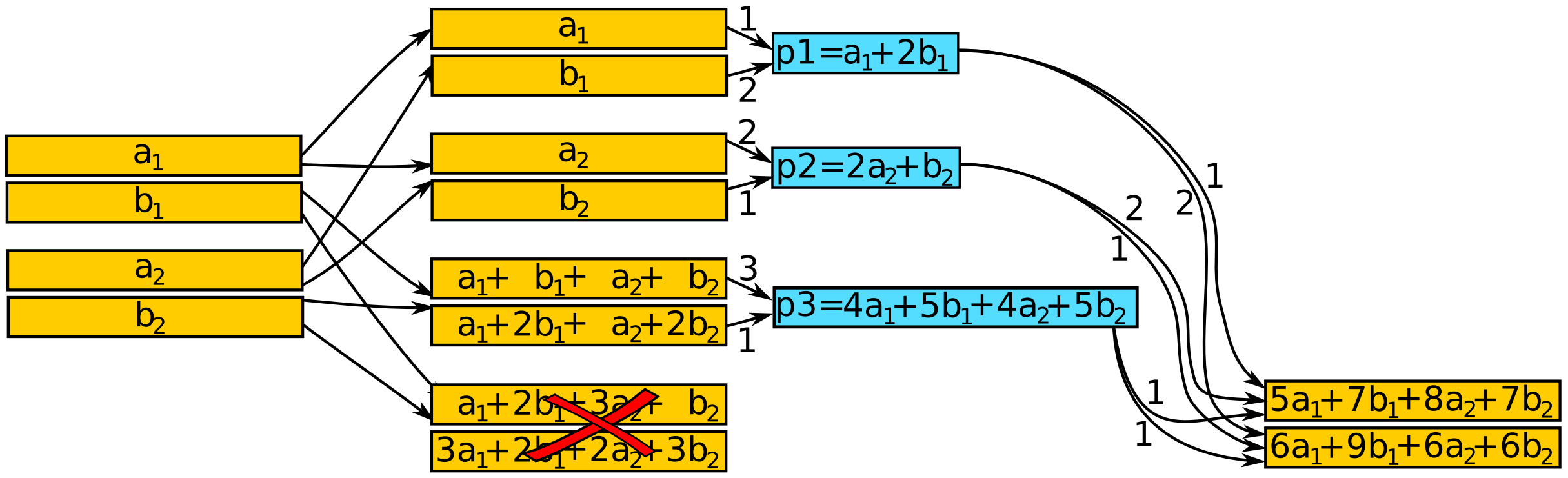

Fig. 2. Example: A repair for a (4,2)-Minimum-Storage Regenerating Code. All the packets (boxes) in this figure have size $0.5$Mb and each node stores two packets. Note that any two nodes have four equations that can be used to recover the data, $a_1, a_2, b_1, b_2$. The parity packets $p_1, p_2, p_3$ are used to create the two packets of the newcomer, requiring repair bandwidth of $1.5$MB. The multiplying coefficients are selected at random and the example is shown over the integers for simplicity (although any sufficiently large field would be enough). The key point is that nodes do not send their information but generate smaller parity packets of their data and forward them to the newcomer who further mixes them to generate two new packets. Note that the selected coefficients also need to be included in the packets, which introduces some overhead.

In this paper we show that, surprisingly, there exist erasure codes that can be repaired without communicating the whole data object. In particular, for the $(4, 2)$ example, we show that the newcomer can download $1.5$Mb to repair a failure and that this is the information theoretic minimum (see Fig. 2 for an example). More generally, we identify a tradeoff between storage and repair bandwidth and show that codes exist that achieve every point on this optimal tradeoff curve. We call codes that lie on this optimal tradeoff curve *regenerating codes*. Note that the tradeoff region computed corrects an error in the threshold $a_c$ computed in [1] and generalizes the result to every feasible $(\alpha, \gamma)$ pair.

The two extremal points on the tradeoff curve are of special interest and we refer to them as minimum-storage regenerating (MSR) codes and minimum-bandwidth regenerating (MBR) codes. The former correspond to Maximum Distance Separable (MDS) codes that can also be efficiently repaired. At the other end of the tradeoff are the MBR codes, which have minimum repair bandwidth. We show that if each storage node is allowed to store slightly more than $\mathcal{M}/k$ bits, the repair bandwidth can be significantly reduced.

The remainder of this paper is organized as follows. In Section II we discuss relevant background and related work from network coding theory and distributed storage systems. In Section III we introduce the notion of the information flow graph, which represents how information is communicated and stored in the network as nodes join and leave the system. In Section III-B we characterize the minimum storage and repair bandwidth and show that there is a tradeoff between these two quantities that can be expressed in terms of a maximum flow on this graph. We further show that for any finite information flow graph, there exists a regenerating code that can achieve any point on the minimum storage/ bandwidth feasible region we computed. Finally, in Section IV we evaluate the performance of the proposed regenerating codes using traces of failures in real systems and compare to alternative schemes previously proposed in the distributed storage literature.

## II. Background and Related Work

### A. Erasure codes

Classical coding theory focuses on the tradeoff between redundancy and error tolerance. In terms of the redundancy-reliability tradeoff, the Maximum Distance Separable (MDS) codes are optimal. The most well-known class of MDS erasure codes is the Reed-Solomon code. More recent studies on erasure coding focus on other performance metrics. For instance, sparse graph codes [11], [12], [13] can achieve near-optimal performance in terms of the redundancy-reliability tradeoff and also require low encoding and decoding complexity. Another line of research for erasure coding in storage applications is parity array codes; see, e.g., [14], [15], [16], [17]. The array codes are based solely on XOR operations and they are generally designed with the objective of low encoding, decoding, and update complexities. Plank [18] gave a tutorial on erasure codes for storage applications at USENIX FAST 2005, which covers Reed-Solomon codes, parity-array codes, and LDPC codes.

Compared to these studies, this paper focuses on different performance metrics. Specifically, motivated by practical concerns in large distributed storage systems, we explore erasure codes that offer good tradeoffs in terms of redundancy, reliability, and repair bandwidth tradeoff.

### B. Network Coding

Network coding is a generalization of the conventional routing (store-and-forwarding) method. In conventional routing, each intermediate node in the network simply stores and forwards information received. In contrast, network coding allows the intermediate nodes to generate output data by encoding (i.e., computing certain functions of) previously received input data. Thus, network coding allows information to be "mixed" at intermediate nodes. The potential advantages of network coding over routing include resource (e.g., bandwidth and power) efficiency, computational efficiency, and robustness to network dynamics. As shown by the pioneering work of Ahlswede et al. [19], network coding can increase the possible network throughput, and in the multicast case can achieve the maximum data rate theoretically possible.

Subsequent work [20], [21] showed that the maximum multicast capacity can be achieved by using linear encoding functions at each node. The studies by Ho *et al.* [22] and Sanders *et al.* [23] further showed that random linear network coding over a sufficiently large finite field can (asymptotically) achieve the multicast capacity. A polynomial complexity procedure to construct deterministic network codes that achieve the multicast capacity is given by Jaggi *et al.* [24].

For distributed storage, the idea of using network coding was introduced in [6] for wireless sensor networks. Many aspects of coding for storage were further explored [7], [25], [26] for sensor network applications. Network coding was proposed for peer-to-peer content distribution systems [27] where random linear operations over packets are performed to improve file downloading in large unstructured overlay networks.

The key difference of this paper to this existing literature is that we bring the dimension of *repair bandwidth* into the picture, and present fundamental bounds and constructions for network codes that need to be maintained over time. Similar to this related work, intermediate nodes form linear combinations in a finite field and the combination coefficients are also stored in each packet, creating some overhead that can be made arbitrarily small for larger packet sizes. In regenerating codes, repair bandwidth is reduced because many nodes create small parity packets of their data that essentially contain enough novel information to generate a new encoded fragment, without requiring to reconstruct the whole data object.

### C. Distributed storage systems

A number of recent studies [28], [8], [29], [30], [4], [31] have designed and evaluated large-scale, peer-to-peer distributed storage systems. Redundancy management strategies for such systems have been evaluated in [9], [32], [4], [10], [31], [33], [34], [35].

Among these, [9], [4], [10] compared replication with erasure codes in the bandwidth-reliability tradeoff space. The analysis of Weatherspoon and Kubiatowicz [9] showed that erasure codes could reduce bandwidth use by an order of magnitude compared with replication. Bhagwan et al. [4] came to a similar conclusion in a simulation of the Total Recall storage system.

Rodrigues and Liskov [10] propose a solution to the repair problem that we call the *Hybrid strategy*: one special storage node maintains one full replica in addition to multiple erasure-coded fragments. The node storing the replica can produce new fragments and send them to newcomers, thus transferring just $\mathcal{M}/k$ bytes for a new fragment. However, maintaining an extra replica on one node dilutes the bandwidth-efficiency of erasure codes and complicates system design. For example, if the replica is lost, new fragments cannot be created until it is restored. The authors show that in high-churn environments (i.e., high rate of node joins/leaves), erasure codes provide a large storage benefits but the bandwidth cost is too high to be practical for a P2P distributed storage system, using the Hybrid strategy. In low-churn environments, the reduction in bandwidth is negligible. In moderate-churn environments, there is some benefit, but this may be outweighed by the added architectural complexity that erasure codes introduce as discussed further in Section IV-E. These conclusions were based on an analytical model augmented with parameters estimated from traces of real systems. Compared with [9], [10] used a much smaller value of $k$ ($7$ instead of $32$) and the Hybrid strategy to address the code regeneration problem. In Section IV, we follow the evaluation methodology of [10] to measure the performance of the two redundancy maintenance schemes that we introduce.

## III. Analysis

Our analysis is based on a particular graphical representation of a distributed storage system, which we refer to as an *information flow graph* $\mathcal{G}$. This graph describes how the information of the data object is communicated through the network, stored in nodes with limited memory, and reaches reconstruction points at the data collectors.

### A. Information Flow Graph

The information flow graph is a directed acyclic graph consisting of three kinds of nodes: a single data source $\mathsf{S}$, storage nodes $\mathsf{x}^i_{in}, \mathsf{x}^i_{out}$ and data collectors $\mathsf{DC}_i$. The single node $\mathsf{S}$ corresponds to the source of the original data. Storage node $i$ in the system is represented by a storage input node $\mathsf{x}^i_{in}$, and a storage output node $\mathsf{x}^i_{out}$; these two nodes are connected by a directed edge $\mathsf{x}^i_{in} \rightarrow \mathsf{x}^i_{out}$ with capacity equal to the amount of data stored at node $i$. See Figure 3 for an illustration.

Given the dynamic nature of the storage systems that we consider, the information flow graph also evolves in time. At any given time, each vertex in the graph is either *active* or *inactive*, depending on whether it is available in the network. At the initial time, only the source node $\mathsf{S}$ is active; it then contacts an initial set of storage nodes, and connects to their inputs ($\mathsf{x}_{in}$) with directed edges of infinite capacity. From this point onwards, the original source node $\mathsf{S}$ becomes and remains inactive. At the next time step, the initially chosen storage nodes become now active; they represent a distributed erasure code, corresponding to the desired steady state of the system. If a new node $j$ joins the system, it can only be connected with active nodes. If the newcomer $j$ chooses to connect with active storage node $i$, then we add a directed edge from $x^i_{out}$ to $x^j_{in}$, with capacity equal to the amount of data that the newcomer downloads from node $i$. Note that in general it is possible for nodes to download more data than they store, as in the example of the $(4, 2)$-erasure code. If a node leaves the system, it becomes inactive. Finally, a data collector DC is a node that corresponds to a request to reconstruct the data. Data collectors connect to subsets of active nodes through edges with infinite capacity.

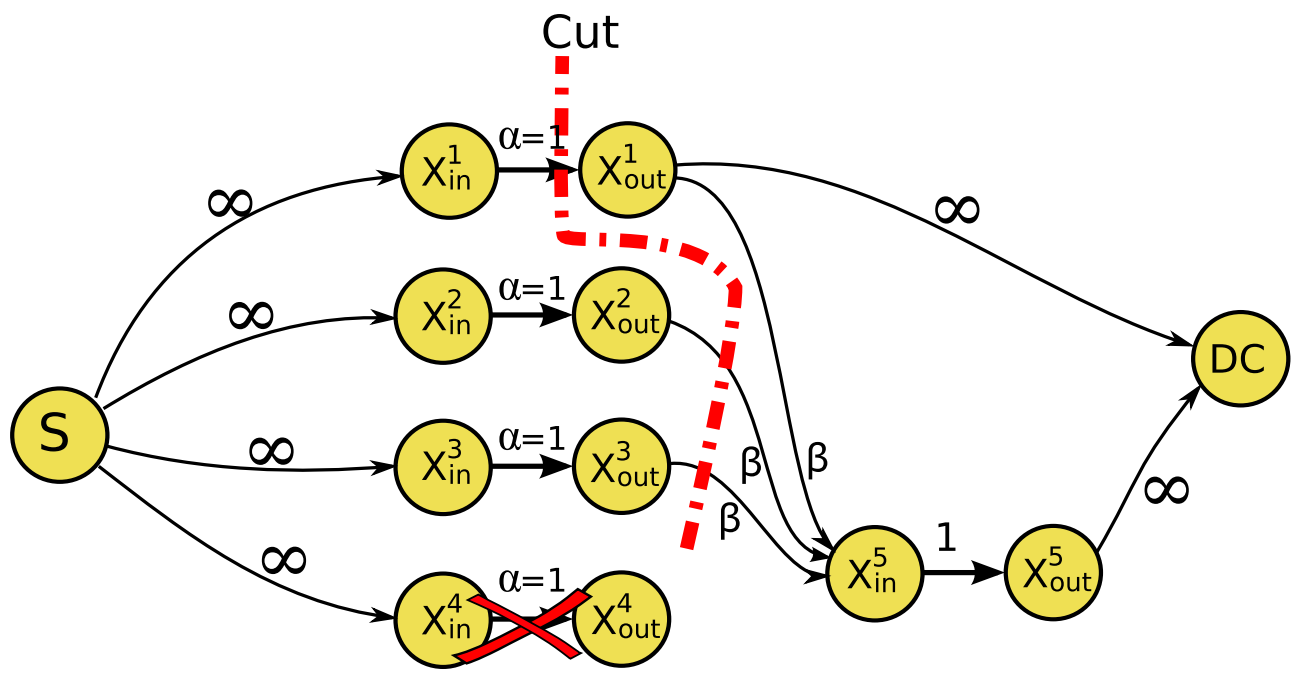

Fig. 3. Illustration of the information flow graph $\mathcal{G}$ corresponding to the (4,2) code of figure 1. A distributed storage scheme uses an $(4, 2)$ erasure code in which any 2 fragments suffice to recover the original data. If node $x^4$ becomes unavailable and a new node joins the system, we need to construct new encoded fragment in $x^5$. To do so, node $x^5_{in}$ is connected to the $d = 3$ active storage nodes. Assuming $\beta$ bits communicated from each active storage node, of interest is the minimum $\beta$ required. The min-cut separating the source and the data collector must be larger than $\mathcal{M} = 2\text{Mb}$ for reconstruction to be possible. For this graph, the min-cut value is given by $1 + 2\beta$, implying that $\beta \geq 0.5\text{Mb}$ is sufficient and necessary.

An important notion associated with the information flow graph is that of minimum cuts: A cut in the graph $\mathcal{G}$ between the source S and a fixed data collector node DC is a subset $C$ of edges such that, there is no path starting from S to DC that does not have one or more edges in $C$. The minimum cut is the cut between S and DC in which the total sum of the edge capacities is smallest.

### *B. Storage-Bandwidth Tradeoff*

We are now ready for the main result of this paper, the characterization of the feasible storage-repair bandwidth points. The setup is as follows: The normal redundancy we want to maintain requires $n$ active storage nodes, each storing $\alpha$ bits. Whenever a node fails, a newcomer downloads $\beta$ bits each from any $d$ surviving nodes. Therefore the total repair bandwidth is $\gamma = d\beta$ (see figure 3). We restrict our attention to the symmetric setup where it is required that any $k$ storage nodes can recover the original file, and a newcomer downloads the same amount of information from each of the existing nodes.

For each set of parameters $(n, k, d, \alpha, \gamma)$, there is a family of information flow graphs, each of which corresponds to a particular evolution of node failures/repairs. We denote this family of directed acyclic graphs by $\mathcal{G}(n, k, d, \alpha, \gamma)$. An $(n, k, d, \alpha, \gamma)$ tuple will be feasible, if a code with storage $\alpha$ and repair bandwidth $\gamma$ exists. For the example in figure 3, the point $(4, 2, 3, 1\text{Mb}, 1.5\text{Mb})$ is feasible (and a code that achieves it is shown in figure 2) and also on the optimal tradeoff whereas a standard erasure code which communicates the whole data object would correspond to $\gamma = 2\text{Mb}$ instead. Note that $n, k, d$ must be integers while $\alpha, \beta, \gamma$ are real valued.

*Theorem 1:* For any $\alpha \geq \alpha^*(d, \gamma)$, the points $(n, k, d, \alpha, \gamma)$ are feasible, and linear network codes suffice to achieve them. It is information theoretically impossible to achieve points with $\alpha < \alpha^*(d, \gamma)$. The threshold function $\alpha^*(d, \gamma)$ (which also depends on $n, k$) is the following:

$$\alpha^*(d,\gamma) = \begin{cases} \frac{\mathcal{M}}{k}, & \gamma \in [f(0), +\infty) \\ \frac{\mathcal{M} - g(i)\gamma}{k-i}, & \gamma \in [f(i), f(i-1)), \end{cases} \tag{1}$$

where

$$f(i) \triangleq \frac{2\mathcal{M}d}{(2k-i-1)i + 2k(d-k+1)}, \tag{2}$$

$$g(i) \triangleq \frac{(2d-2k+i+1)i}{2d}. \tag{3}$$

The minimum $\gamma$ is

$$\gamma_{\min} = f(k-1) = \frac{2\mathcal{M}d}{2kd - k^2 + k}. \tag{4}$$

The complete proof of this theorem is given in the Appendix. The main idea is that the code repair problem can be mapped to a multicasting problem on the information flow graph. Known results on network coding for multicasting can then be used to establish that code repair can be achieved if and only if the underlying information flow graph has enough connectivity. The bulk of the technical analysis of the proof then involves computing the minimum cuts on arbitrary graphs in $\mathcal{G}(n, k, d, \alpha, \gamma)$ and solving an optimization problem for minimizing $\alpha$ subject to a sufficient flow constraint.

The optimal tradeoff curves for $k = 5, n = 10, d = 9$ and $k = 10, n = 15, d = 14$ are shown in Figure 4 (top) and (bottom), respectively.

### *C. Special Cases: Minimum-Storage Regenerating (MSR) Codes and Minimum-Bandwidth Regenerating (MBR) Codes*

We now study two extremal points on the optimal tradeoff curve, which correspond to the best storage efficiency and the minimum repair bandwidth, respectively. We call codes that attain these points minimum-storage regenerating (MSR) codes and minimum-bandwidth regenerating (MBR) codes, respectively.

It can be verified that the minimum storage point is achieved by the pair

$$(\alpha_{MSR}, \gamma_{MSR}) = \left(\frac{\mathcal{M}}{k}, \frac{\mathcal{M}d}{k(d-k+1)}\right). \tag{5}$$

If we substitute $d = k$ into the above, we note that the total network bandwidth for repair is $\mathcal{M}$, the size of the original file. Therefore, if we only allow a newcomer to contact $k$ nodes, it is optimal to download the whole file and then compute the new fragment. However, if we allow a newcomer to contact more than $k$ nodes, the network bandwidth $\gamma_{MSR}$ can be reduced significantly. The minimum network bandwidth is clearly achieved by having the newcomer contact all other nodes. For instance, for $(n, k) = (14, 7)$, the newcomer needs to download only $\frac{\mathcal{M}}{49}$ from each of the $d = n - 1 = 13$ active storage nodes, making the repair bandwidth equal to $\frac{13\mathcal{M}}{49}$, required to generate a fragment of size $\frac{\mathcal{M}}{7}$.

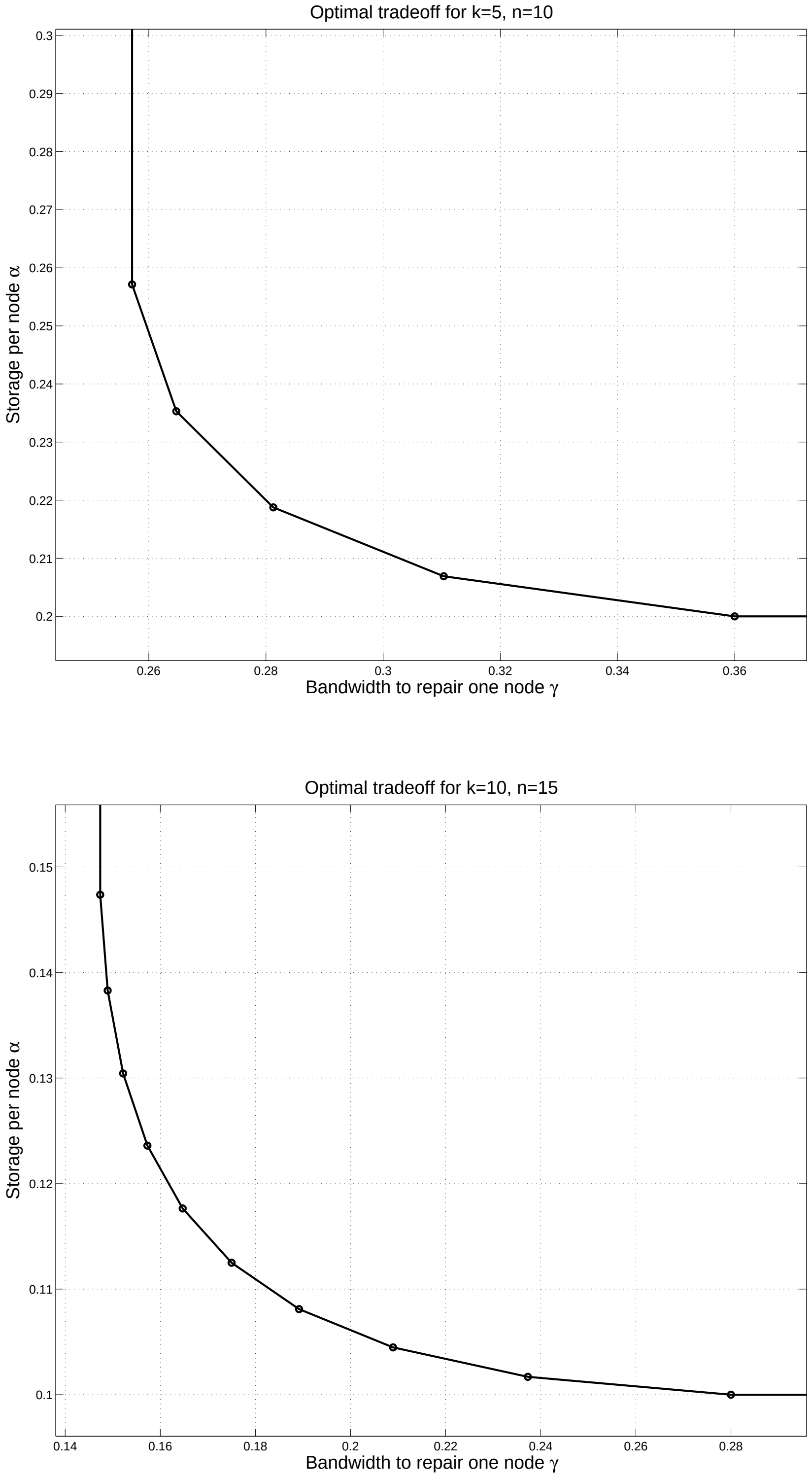

Fig. 4. Optimal tradeoff curve between storage $\alpha$ and repair bandwidth $\gamma$, for $k = 5, n = 10$ (left) and $k = 10, n = 15$ (right). For both plots $\mathcal{M} = 1$ and $d = n - 1$. Note that traditional erasure coding corresponds to the points $(\gamma = 1, \alpha = 0.2)$ and $(\gamma = 1, \alpha = 0.1)$ for the top and bottom plots.

Since the MSR codes store $\frac{\mathcal{M}}{k}$ bits at each node while ensuring any $k$ coded blocks can be used to recover the original file, the MSR codes have equivalent reliability-redundancy performance with standard Maximum Distance Separable (MDS) codes. However, MSR codes outperform classical MDS codes in terms of the network repair bandwidth.

At the other end of the tradeoff are MBR codes, which have minimum repair bandwidth. It can be verified that the minimum repair bandwidth point is achieved by

$$(\alpha_{MBR}, \gamma_{MBR}) = \left( \frac{2\mathcal{M}d}{2kd - k^2 + k}, \ \frac{2\mathcal{M}d}{2kd - k^2 + k} \right). \quad (6)$$

Note that the minimum bandwidth regenerating codes, the storage size $\alpha$ is equal to $\gamma$, the total number of bits downloaded. Therefore MBR codes incur no bandwidth expansion at all, just like a replication system does. However, the benefit of MBR codes is significantly better storage efficiency.

## IV. Evaluation

In this section, we compare regenerating codes with other redundancy management schemes in the context of distributed storage systems. We follow the evaluation methodology of [10], which consists of a simple analytical model whose parameters are obtained from traces of node availability measured in several real distributed systems.

We begin in Section IV-A with a discussion of node dynamics and the objectives relevant to distributed storage systems, namely reliability, bandwidth, and disk space. We introduce the model in Section IV-B and estimate realistic values for its parameters in Section IV-C. Section IV-D contains the quantitative results of our evaluation. In Section IV-E, we discuss qualitative tradeoffs between regenerating codes and other strategies, and how our results change the conclusion of [10] that erasure codes provide limited practical benefit.

### A. Node dynamics and objectives

In this section we introduce some background and terminology which is common to most of the work discussed in Section II-C.

We draw a distinction between *permanent* and *transient* node failures. A permanent failure, such as the permanent departure of a node from the system or a disk failure, results in loss of the data stored on the node. In contrast, data is preserved across a transient failure, such as a reboot or temporary network disconnection. We say that a node is *available* when its data can be retrieved across the network.

Distributed storage systems attempt to provide two types of reliability: availability and durability. A file is *available* when it can be reconstructed from the data stored on currently available nodes. A file's *durability* is maintained if it has not been lost due to permanent node failures: that is, it may be available at some point in the future. Both properties are desirable, but in this paper we report results for availability only. Specifically, we will show *file unavailability*, the fraction of time that the file is not available.

### B. Model

We use a model which is intended to capture the average-case bandwidth used to maintain a file in the system, and the resulting average availability of the file. With minor exceptions,[1]this model and the subsequent estimation of its parameters are equivalent to that of [10]. Although this evaluation methodology is a significant simplification of real storage systems, it allows us to compare directly with the conclusions of [10] as well as to calculate precise values for rare events.

The model has two key parameters, $f$ and $a$. First, we assume that in expectation a fraction $f$ of the nodes storing file data fail permanently per unit time, causing data transfers to repair the lost redundancy. Second, we assume that at any given time while a node is storing data, the node is available with some probability $a$ (and with probability $1-a$ is currently experiencing a transient failure). Moreover, the model assumes that the event that a node is available is independent of the availability of all other nodes.

Under these assumptions, we can compute the expected availability and maintenance bandwidth of various redundancy schemes to maintain a file of $\mathcal{M}$ bytes. We make use of the fact that for all schemes except MSR codes, the amount of bandwidth used is equal to the amount of redundancy that had to be replaced, which is in expectation $f$ times the amount of storage used.

**Replication:** If we store $\mathcal{R}$ replicas of the file, then we store a total of $\mathcal{R} \cdot \mathcal{M}$ bytes, and in expectation we must replace $f \cdot \mathcal{R} \cdot \mathcal{M}$ bytes per unit time. The file is unavailable if no replica is available, which happens with probability $(1-a)^{\mathcal{R}}$.

**Ideal Erasure Codes:** For comparison, we show the bandwidth and availability of a hypothetical $(n, k)$ erasure code strategy which can "magically" create a new packet while transferring just $\mathcal{M}/k$ bytes (*i.e.*, the size of the packet). Setting $n = k \cdot \mathcal{R}$, this strategy sends $f \cdot \mathcal{R} \cdot \mathcal{M}$ bytes per unit time and has unavailability probability $U_{\text{ideal}}(n, k) := \sum_{i=0}^{k-1} \binom{n}{i} a^i (1-a)^{n-i}$.

**Hybrid:** If we store one full replica plus an $(n, k)$ erasure code where $n = k \cdot (\mathcal{R} - 1)$, then we again store $\mathcal{R} \cdot \mathcal{M}$ bytes in total, so we transfer $f \cdot \mathcal{R} \cdot \mathcal{M}$ bytes per unit time in expectation. The file is unavailable if the replica is unavailable *and* fewer than $k$ erasure-coded packets are available, which happens with probability $(1-a) \cdot U_{\text{ideal}}(n, k)$.

**Minimum-Storage Regenerating Codes:** An $(n, k)$ MSR Code with redundancy $\mathcal{R} = n/k$ stores $\mathcal{R}\mathcal{M}$ bytes in total, so $f \cdot \mathcal{R} \cdot \mathcal{M}$ bytes must be replaced per unit time. We will refer to the *overhead* of an MSR code $\delta_{MSR}$ as the extra amount

[1]In addition to evaluating a larger set of strategies and using a somewhat different set of traces, we count bandwidth cost due to permanent node failure only, rather than both failures and joins. Most designs [4], [31], [33] can avoid reacting to node joins. Additionally, we compute probabilities directly rather than using approximations to the binomial.

of information that needs to be transfered compared to the fragment size $\mathcal{M}/k$:

$$\delta_{MSR} \triangleq \frac{(n-1)\beta_{MSR}}{\mathcal{M}/k} = \frac{n-1}{n-k}. \quad (7)$$

Therefore, replacing a fragment requires transferring over the network $\delta_{\text{MSR}}$ times the size of the fragment in the most favorable case when newcomers connect to $d = n-1$ nodes to construct a new fragment. Therefore, this results in $f \cdot \mathcal{R} \cdot \mathcal{M} \cdot \delta_{\text{MSR}}$ bytes sent per unit time, and unavailability $U_{\text{ideal}}(n,k)$.

**Minimum-Bandwidth Regenerating Codes:**

It is convenient to define the MBR code overhead as the amount of information transfered over the ideal fragment size:

$$\delta_{MBR} \triangleq \frac{(n-1)\beta_{MBR}}{\mathcal{M}/k} = \frac{2(n-1)}{2n-k-1}. \quad (8)$$

Therefore, an $(n,k)$ MBR Code stores $\mathcal{M} \cdot n \cdot \delta_{\text{MBR}}$ bytes in total. So in expectation $f \cdot \mathcal{M} \cdot n \cdot \delta_{\text{MBR}}$ bytes are transfered per unit time, and the unavailability is again $U_{\text{ideal}}(n,k)$.

### C. Estimating $f$ and $a$

In this section we describe how we estimate $f$, the fraction of nodes that permanently fail per unit time, and $a$, the mean node availability, based on traces of node availability in several distributed systems.

We use four traces of node availability with widely varying characteristics, summarized in Table I. The **PlanetLab All Pairs Ping [36]** trace is based on pings sent every 15 minutes between all pairs of 200-400 nodes in PlanetLab, a stable, managed network research testbed. We consider a node to be up in one 15-minute interval when at least half of the pings sent to it in that interval succeeded. In a number of periods, all or nearly all PlanetLab nodes were down, most likely due to planned system upgrades or measurement errors. To exclude these cases, we "cleaned" the trace as follows: for each period of downtime at a particular node, we remove that period (i.e. we consider the node up during that interval) when the average number of nodes up during that period is less than half the average number of nodes up over all time. The **Microsoft PCs [28]** trace is derived from hourly pings to desktop PCs within Microsoft Corporation. The **Skype superpeers [37]** trace is based on application-level pings at 30-minute intervals to nodes in the Skype superpeer network, which may approximate the behavior of a set of well-provisioned endhosts, since superpeers may be selected in part based on bandwidth availability [37]. Finally, the trace of **Gnutella peers [38]** is based on application-level pings to ordinary Gnutella peers at 7-minute intervals.

We next describe how we derive $f$ and $a$ from these traces. It is of key importance for the storage system to distinguish between permanent and transient failures (defined in Section IV-A), since only the former requires bandwidth-intensive replacement of lost redundancy. Most systems use a *timeout* heuristic: when a node has not responded to network-level probes after some period of time $t$, it is considered to have failed permanently. To approximate a storage system's behavior, we use the same heuristic. Node availability $a$ is then calculated as the mean (over time) fraction of nodes which were available among those which were not considered permanently failed at that time.

The resulting values of $f$ and $a$ appear in Table I, where we have fixed the timeout $t$ at $1$ day. Longer timeouts reduce overall bandwidth costs [10], [33], but begin to impact durability [33] and are more likely to produce artificial effects in the short ($2.5$-day) Gnutella trace.

We emphasize that the procedure described above only provides an estimate of $f$ and $a$ which may be biased in several ways. Some designs [33] reincorporate data on nodes which return after transient failures which were longer than the timeout $t$, which would reduce $f$. Additionally, even placing files on uniform-random nodes results in selecting nodes that are more available [34] and less prone to failure [35] than the average node. Finally, we have not accounted for the time needed to transfer data onto a node, during which it is effectively unavailable. However, we consider it unlikely that these biases would impact our main results since we are primarily concerned with the *relative* performance of the strategies we compare.

### D. Quantitative results

Figure 5 shows the tradeoff between mean unavailability and mean maintenance bandwidth in each of the strategies of Section IV-B using the values of $f$ and $a$ from Section IV-C and $k = 7$. Feasible points in the tradeoff space are produced by varying the redundancy factor $\mathcal{R}$. The marked points along each curve highlight a subset of the feasible points (i.e., points for which $n$ is integral).

Figure 6 shows that relative performance of the various strategies is similar for $k = 14$.

For conciseness, we omit plots of storage used by the schemes. However, disk usage is proportional to bandwidth for all schemes we evaluate in this section, with the exception of minimum storage regenerating codes. This is because MSR codes are the only scheme in which the data transferred onto a newcomer is not equal to the amount of data that the newcomer finally stores. Instead, the storage used by MSR codes is equal to that of the storage used by hypothetical ideal erasure codes, and hence MSR codes' space usage is proportional to the bandwidth used by ideal codes.

For example, from Figure 5(b) we can compare the strategies at their feasible points closest to unavailability $0.0001$, i.e., four nines of availability. At these points, MSR codes use about $44\%$ more bandwidth and $28\%$ less storage space than Hybrid, while MBR codes use about $3.7\%$ less bandwidth and storage space than Hybrid. Additionally, these feasible points give MSR and MBR codes somewhat better unavailability than Hybrid ($.000059$ vs. $0.00018$).

One interesting effect apparent in the plots is that MSR codes' maintenance bandwidth actually *decreases* as the redundancy factor $\mathcal{R}$ increases, before coming to a minimum and then increasing again. Intuitively, while increasing $\mathcal{R}$ increases the total amount of data that needs to be maintained, for small $\mathcal{R}$ this is more than compensated for by the reduction in overhead. The expected maintenance bandwidth per unit time

| **Trace** | **Length (days)** | **Start date** | **Mean # nodes up** | $f$ **(fraction failed per day)** | $a$ |
|---|---|---|---|---|---|
| PlanetLab | 527 | Jan. 2004 | 303 | 0.017 | 0.97 |
| Microsoft PCs | 35 | Jul. 6, 1999 | 41970 | 0.038 | 0.91 |
| Skype | 25 | Sept. 12, 2005 | 710 | 0.12 | 0.65 |
| Gnutella | 2.5 | May, 2001 | 1846 | 0.30 | 0.38 |

TABLE I
THE AVAILABILITY TRACES USED IN THIS PAPER.

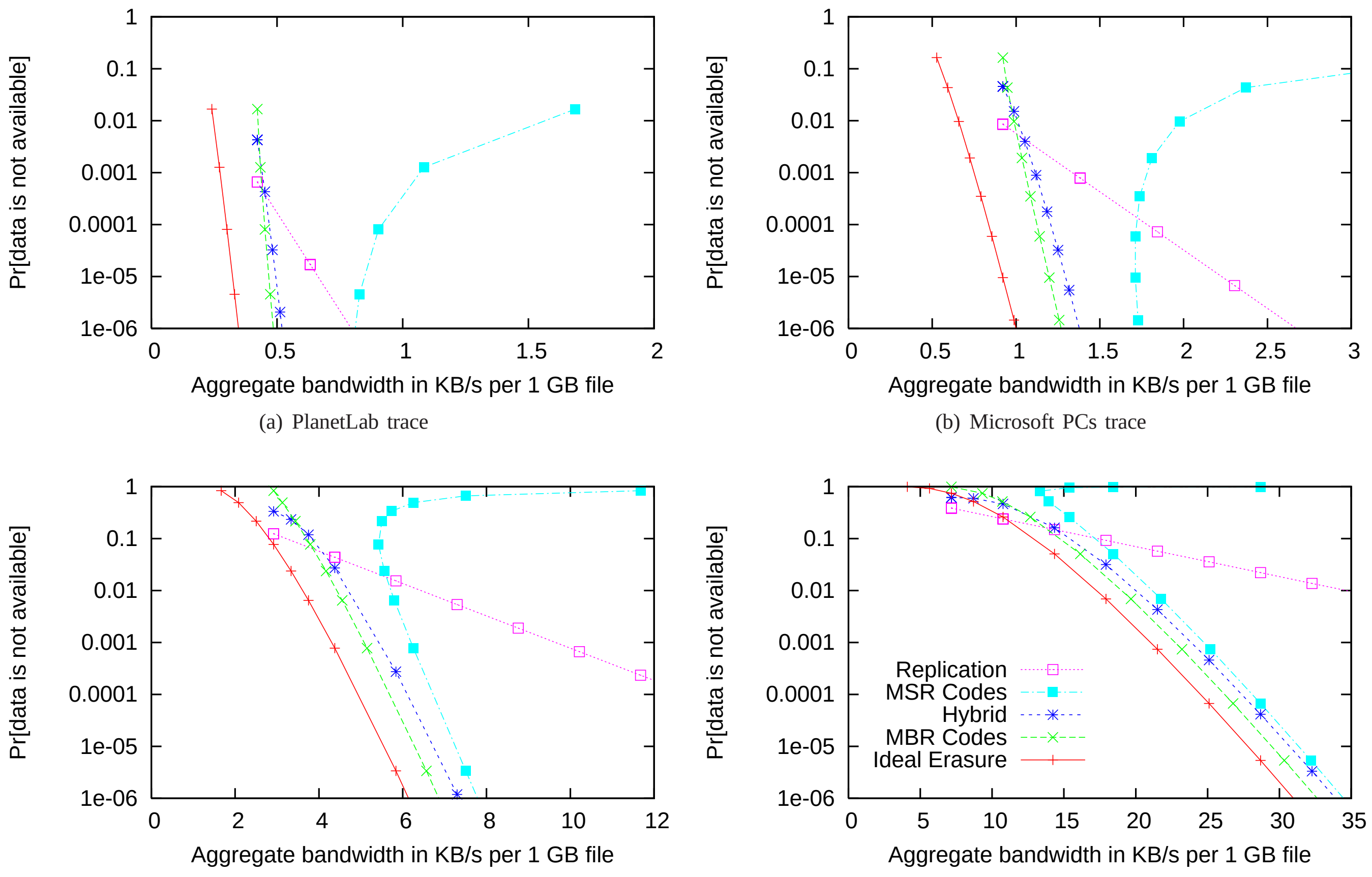

(a) PlanetLab trace

(b) Microsoft PCs trace

(c) Skype superpeers trace

(d) Gnutella peers trace

Fig. 5. Availability-bandwidth tradeoff for $k = 7$ with parameters derived from each of the traces. The key in (d) applies to all four plots.

is

$$f\,\mathcal{M}\,\mathcal{R}\,\delta_{\text{MSR}} = f\,\mathcal{M}\frac{n}{k}\frac{n-1}{n-k}. \quad (9)$$

It is easy to see that this function is minimized by selecting n one of the two integers closest to

$$n_{opt} = k + \sqrt{k^2 - k}. \quad (10)$$

which approaches a redundancy factor of 2 as $k \to \infty$.

## *E. Qualitative comparison*

In this section we discuss two questions: First, based on the results of the previous section, what are the qualitative advantages and disadvantages of the two extremal regenerating codes compared with the Hybrid coding scheme? Second, do our results affect the conclusion of Rodrigues and Liskov [10] that erasure codes offer too little improvement in bandwidth use to clearly offset the added complexity that they add to the system?

*1) Comparison with Hybrid:* Compared with Hybrid, for a given target availability, minimum storage regenerating codes offer slightly lower maintenance bandwidth and storage, and a simpler system architecture since only one type of redundancy needs to be maintained. An important practical disadvantage of using the Hybrid scheme is asymmetric design which can cause the disk I/O to become the bottleneck of the system during repairs. This is because the disc storing the full replica and generates the encoded fragments need to read the whole data object and compute the encoded fragment.

However, MBR codes have at least two disadvantages. First, constructing a new packet, or reconstructing the entire file, requires communcation with $n-1$ nodes[2] rather than one (in Hybrid, the node holding the single replica). This adds overhead that could be significant for sufficiently small files or sufficiently large $n$. Perhaps more importantly, there is a factor $\delta_{\text{MBR}}$ increase in total data transferred to *read* the file, roughly 30% for a redundancy factor $\mathcal{R} = 2$ and $k = 7$ or 13% for $\mathcal{R} = 4$, Thus, if the frequency that a file is read is sufficiently high and $k$ is sufficiently small, this inefficiency could become unacceptable. Again compared with Hybrid, MSR codes offer a simpler, symmetric system design and somewhat lower storage space for the same reliability. However, MSR codes have somewhat higher maintenance bandwidth and like MSB codes require that newcomers and data collectors connect to multiple nodes.

[2]The scheme could be adapted to connect to fewer than $n-1$ nodes, but this would increase maintenance bandwidth.

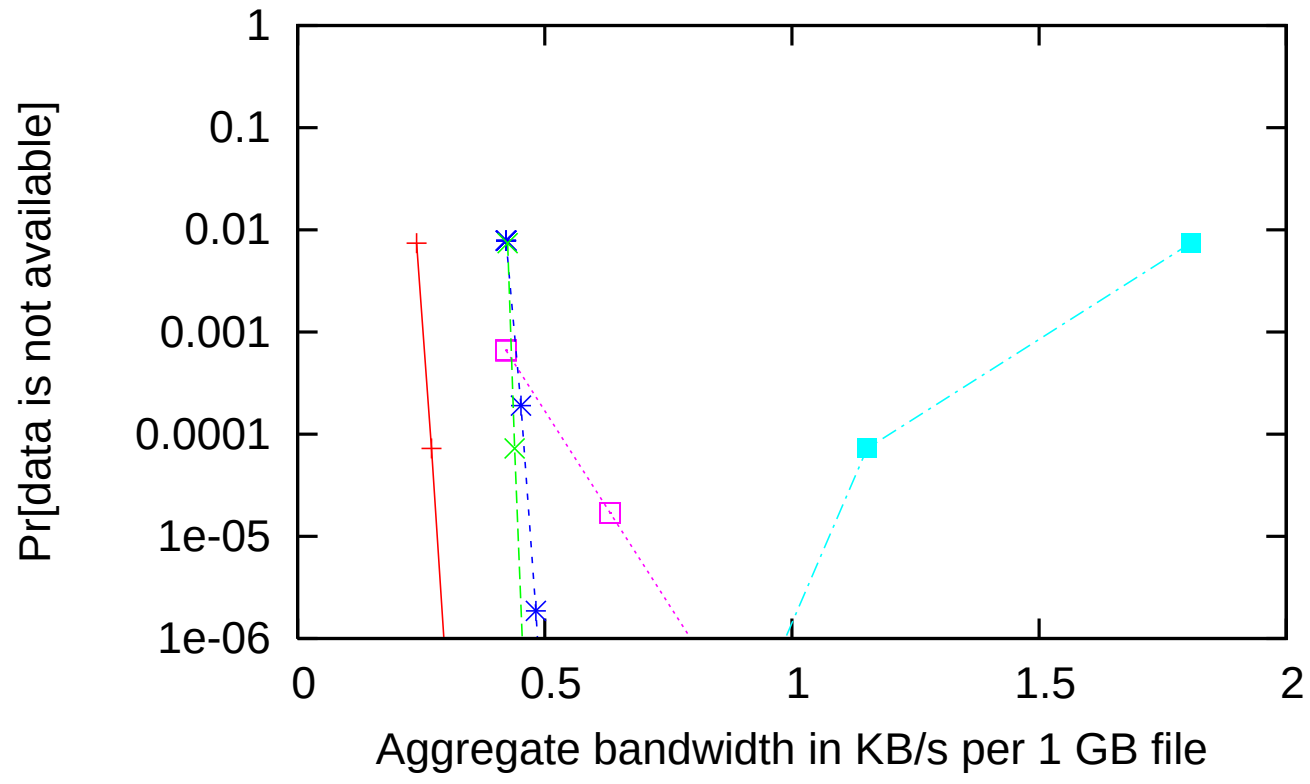

(a) PlanetLab trace

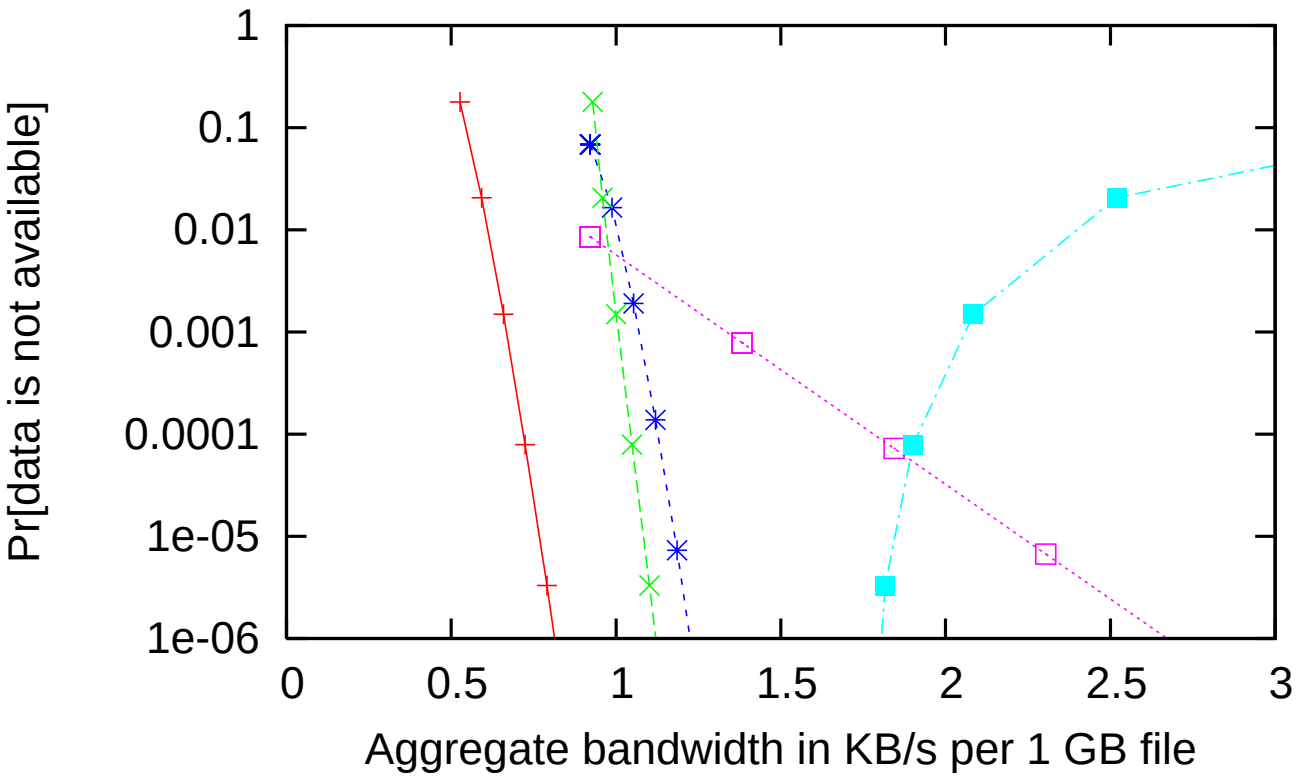

(b) Microsoft PCs trace

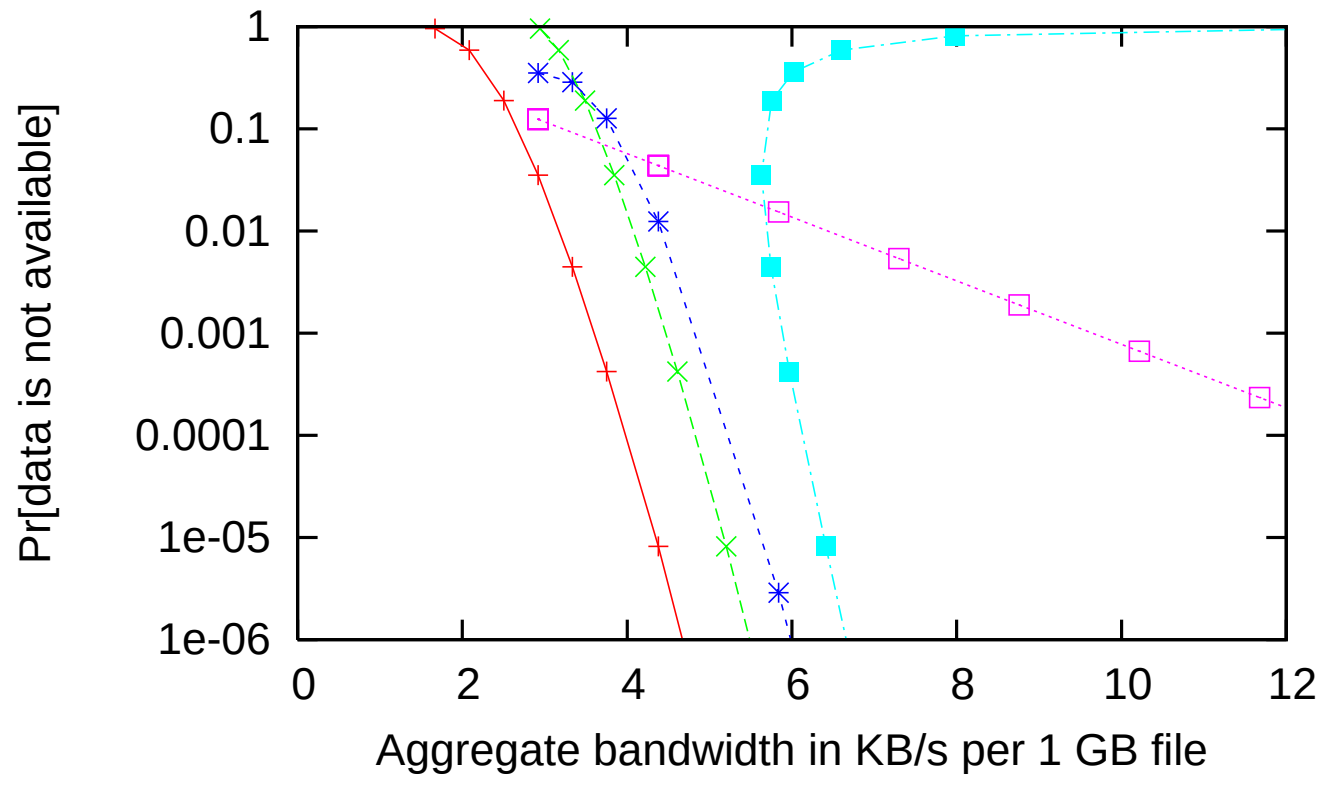

(c) Skype superpeers trace

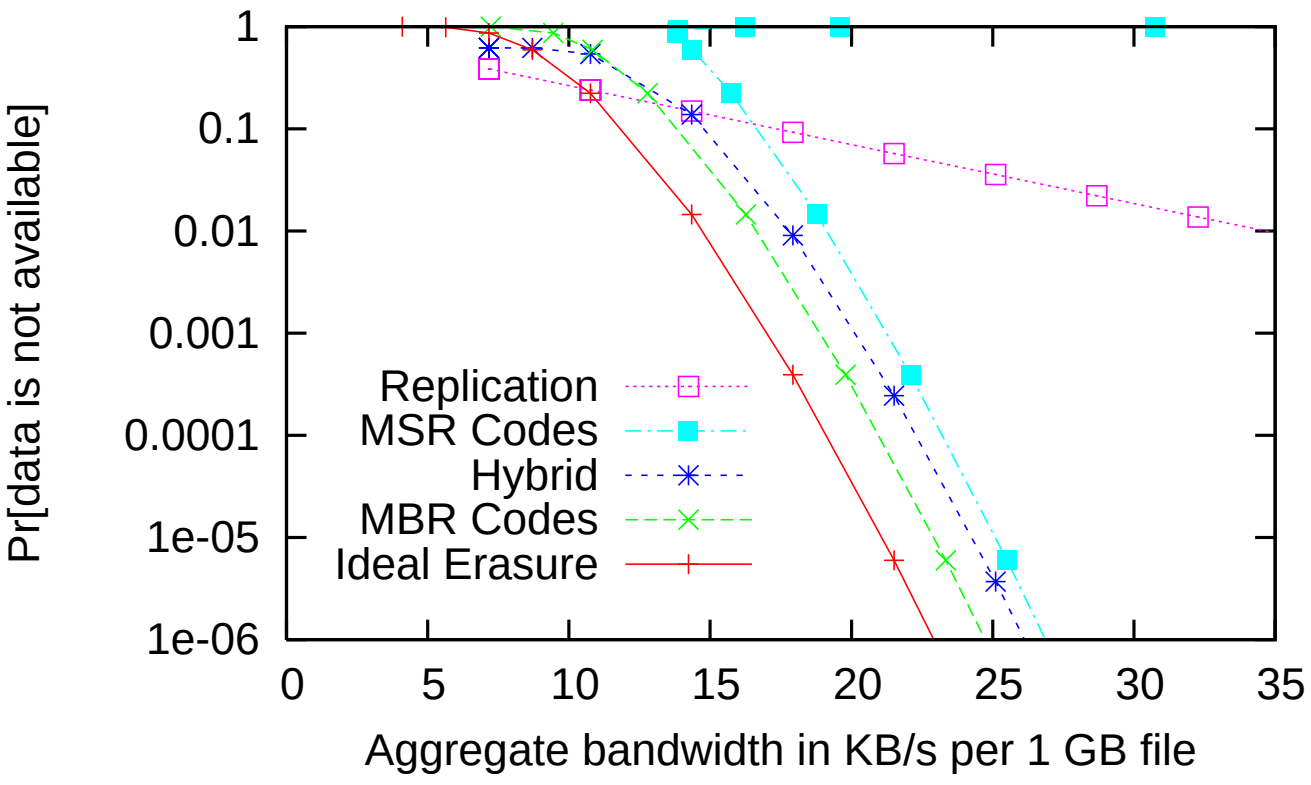

(d) Gnutella peers trace

Fig. 6. Availability-bandwidth tradeoff for $k = 14$ with parameters derived from each of the traces.

Rodrigues et al. [10] discussed two principal disadvantages of using erasure codes in a widely distributed system: coding—in particular, the Hybrid strategy—complicates the system architecture; and the improvement in maintenance bandwidth was minimal in more stable environments, which are the more likely deployment scenario. Regenerating codes address the first of these issues, which may make coding more broadly applicable.

## V. Conclusions

We presented a general theoretic framework that can determine the information that must be communicated to repair failures in encoded systems and identified a tradeoff between storage and repair bandwidth.

Certainly there are many issues that remain to be addressed before these ideas can be implemented in practical systems. In future work we plan to investigate deterministic designs of regenerating codes over small finite fields, the existence of *systematic* regenerating codes, designs that minimize the overhead storage of the coefficients, as well as the impact of node dynamics in reliability. Other issues of interest involve how CPU processing and disk I/O will influence the system performance, as well as integrity and security for the linear combination packets (see [39] for a related analysis for content distribution).

One potential application for the proposed regenerating codes is distributed archival storage or backup, which might be useful for data center applications. In this case, files are likely to be large and infrequently read, making the drawbacks mentioned above less significant, so that MBR codes' symmetric design may make them a win over Hybrid; and the required reliability may also be high, making them a win over simple replication. In other applications (such as storage system within fast local networks) the required storage may become important, and the results of the previous section show that minimum storage regenerating codes can be useful.

## VI. Appendix

Here we prove Theorem 1. We first start with the following simple lemma.

*Lemma 1:* No data collector DC can reconstruct the initial data object if the minimum cut in $\mathcal{G}$ between S and DC is smaller than the initial object size $\mathcal{M}$.

*Proof:* The information of the initial data object must be communicated from the source to the particular data collector. Since every link in the information flow graph can only be used at most once, and since the point-to-point capacity is less than the data object size, a standard cut-set bound shows that the entropy of the data object conditioned on everything observable to the data collector is non-zero and therefore reconstruction is impossible. ∎

The information flow graph casts the original storage problem as a network communication problem where the source $s$ multicasts the file to the set of all possible data collectors. By analyzing the connectivity in the information flow graph, we obtain necessary conditions for all possible storage codes, as shown in Lemma 1. In addition to providing necessary conditions for all codes, the information flow graph can also imply the existence of codes under proper assumptions.

*Proposition 1:* Consider any given finite information flow graph $\mathcal{G}$, with a finite set of data collectors. If the minimum of the min-cuts separating the source with each data collector is larger or equal to the data object size $\mathcal{M}$, then there exists a linear network code defined over a sufficiently large finite field $\mathbb{F}$ (whose size depends on the graph size) such that all data collectors can recover the data object. Further, randomized network coding guarantees that all collectors can recover the data object with probability that can be driven arbitrarily high by increasing the field size.

*Proof:* The key point is observing that the reconstruction problem reduces exactly to multicasting on all the possible data collectors on the information flow graph $\mathcal{G}$. Therefore, the result follows directly from the constructive results in network coding theory for single source multicasting; see the discussion of related works on network coding in Section II-B. ∎

To apply Proposition 1, consider an information flow graph $\mathcal{G}$ that enumerates all possible failure/repair patterns and all possible data collectors when the number of failures/repairs is bounded. This implies that there exists a valid regenerating code achieving the necessary cut bound (cf. Lemma 1), which can tolerate a bounded number of failures/repairs. In another paper [2], we present coding methods that construct deterministic regenerating codes that can tolerate infinite number of failures/repairs, with a bounded field size, assuming only the population of active nodes at any time is bounded. For the detailed coding theoretic construction, please refer to [2].

We analyze the connectivity in the information flow graph to find the minimum repair bandwidth. The next key lemma characterizes the flow in any information flow graph, under arbitrary failure pattern and connectivity.

*Lemma 2:* Consider any (potentially infinite) information flow graph $G$, formed by having $n$ initial nodes that connect directly to the source and obtain $\alpha$ bits, while additional nodes join the graph by connecting to $d$ existing nodes and obtaining $\beta$ bits from each.[3] Any data collector $t$ that connects to a $k$-subset of "out-nodes" (c.f. Figure 3) of $G$ must satisfy:

$$\text{mincut}(s,t) \geq \sum_{i=0}^{\min\{d,k\}-1} \min\{(d-i)\beta, \alpha\}. \qquad (11)$$

Furthermore, there exists an information flow graph $G^* \in \mathcal{G}(n,k,d,\alpha,\beta)$ where this bound is matched with equality.

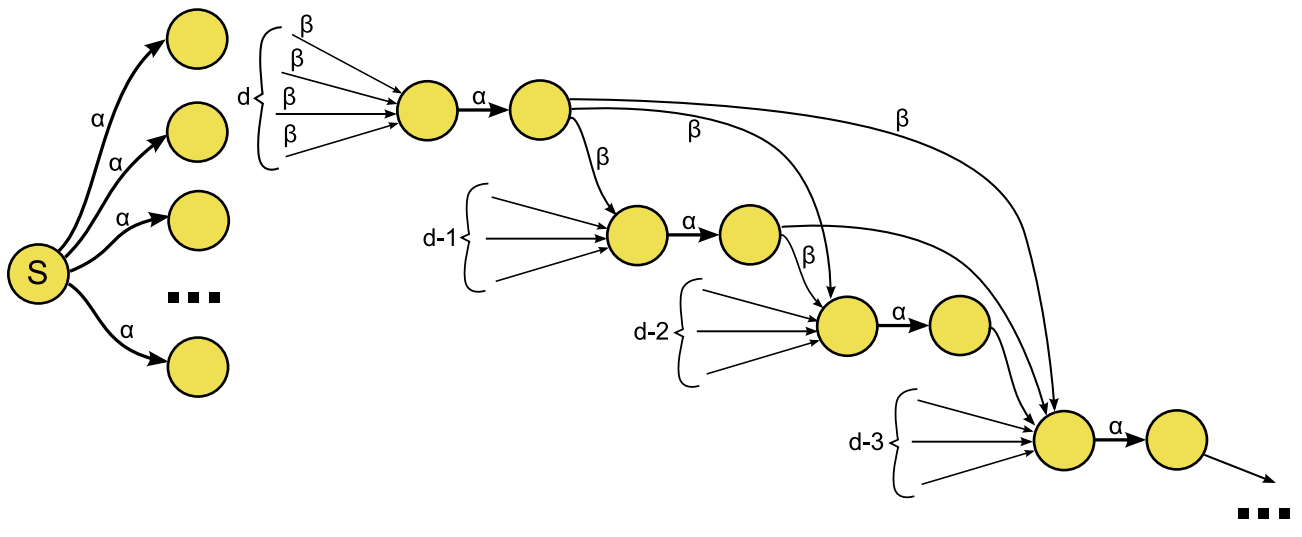

Fig. 7. $G^*$ used in the proof of lemma 2

**Proof:** First, we show that there exists an information flow graph $G^*$ where the bound (11) is matched with equality. This graph is illustrated by Figure 7. In this graph, there are initially $n$ nodes labeled from 1 to $n$. Consider $k$ newcomers labeled as $n+1, \ldots, n+k$. The newcomer node $n+i$ connects to nodes $n+i-d, \ldots, n+i-1$. Consider a data collector $t$ that connects to the last $k$ nodes, i.e., nodes $n+1, \ldots, n+k$. Consider a cut $(U, \overline{U})$ defined as follows. For each $i \in \{1, \ldots, k\}$, if $\alpha \leq (d-i)\beta$, then we include $\mathsf{x}_{out}^{n+i}$ in $\overline{U}$; otherwise, we include $\mathsf{x}_{out}^{n+i}$ and $\mathsf{x}_{in}^{n+i}$ in $\overline{U}$. Then this cut $(U, \overline{U})$ achieves (11) with equality.

We now show that (11) must be satisfied for any $G$ formed by adding $d$ in-degree nodes as described above. Consider a data collector $t$ that connects to a $k$-subset of "out-nodes", say $\{\mathsf{x}_{out}^i : i \in I\}$. We want to show that any $s$–$t$ cut in $G$ has capacity at least

$$\sum_{i=0}^{\min\{d,k\}-1} \min\{(d-i)\beta, \alpha\}. \qquad (12)$$

Since the incoming edges of $t$ all have infinite capacity, we only need to examine the cuts $(U, \overline{U})$ with $s \in U$,

$$\mathsf{x}_{out}^i \in \overline{U}, \forall i \in I. \qquad (13)$$

Let $\mathcal{C}$ denote the edges in the cut, i.e., the set of edges going from $U$ to $\overline{U}$.

Every directed acyclic graph has a topological sorting (see, e.g., [40]), where a topological sorting (or acyclic ordering) is an ordering of its vertices such that the existence of an edge from $v_i$ to $v_j$ implies $i < j$. Let $\mathsf{x}_{out}^1$ be the topologically first output node in $\overline{U}$. Consider two cases:

- If $\mathsf{x}_{in}^1 \in U$, then the edge $\mathsf{x}_{in}^1 \mathsf{x}_{out}^1$ must be in $\mathcal{C}$.
- If $\mathsf{x}_{in}^1 \in \overline{U}$, since $\mathsf{x}_{in}^1$ has an in-degree of $d$ and it is the topologically first node in $\overline{U}$, all the incoming edges of $\mathsf{x}_{in}^1$ must be in $\mathcal{C}$.

Therefore, these edges related to $\mathsf{x}_{out}^1$ will contribute a value of $\min\{d\beta, \alpha\}$ to the cut capacity.

Now consider $\mathsf{x}_{out}^2$, the topologically second output node in $\overline{U}$. Similar to the above, we have two cases:

- If $\mathsf{x}_{in}^2 \in U$, then the edge $\mathsf{x}_{in}^2 \mathsf{x}_{out}^2$ must be in $\mathcal{C}$.
- If $\mathsf{x}_{in}^2 \in \overline{U}$, since at most one of the incoming edges of $\mathsf{x}_{in}^2$ can be from $\mathsf{x}_{out}^1$, $d-1$ incoming edges of $\mathsf{x}_{in}^1$ must be in $\mathcal{C}$.

Following the same reasoning we find that for the $i$-th node ($i = 0, \ldots, \min\{d,k\}-1$) in the sorted set $\overline{U}$, either one edge of capacity $\alpha$ or $(d-i)$ edges of capacity $\beta$ must be in $\mathcal{C}$. Equation (11) is exactly summing these contributions. ∎

From Lemma 2, we know that there exists a graph $G^* \in \mathcal{G}(n,k,d,\alpha,\beta)$ whose mincut is exactly $\sum_{i=0}^{\min\{d,k\}-1} \min\{(d-i)\beta, \alpha\}$. This implies that if we want to ensure recoverability while allowing a newcomer to connect to *any* set of $d$ existing nodes, then the following is a necessary condition[4]

$$\sum_{i=0}^{\min\{d,k\}-1} \min\{(d-i)\beta, \alpha\} \geq \mathcal{M}. \qquad (14)$$

Furthermore, when this condition is satisfied, we know any graph in $\mathcal{G}(n,k,d,\alpha,\beta)$ will have enough flow from the source to each data collector. For this reason, we say

$$\mathsf{C} \triangleq \sum_{i=0}^{\min\{d,k\}-1} \min\{(d-i)\beta, \alpha\} \qquad (15)$$

[3]Note that this setup allows more graphs than those in $\mathcal{G}(n,k,d,\alpha,\beta)$. In a graph in $\mathcal{G}(n,k,d,\alpha,\beta)$, at any time there are $n$ active storage nodes and a newcomer can only connect to the active nodes. In contrast, in a graph $G$ described in this lemma, there is no notion of "active nodes" and a newcomer can connect to any $d$ existing nodes.

[4]This, however, does not rule out the possibility that the mincut is larger if a newcomer can choose the $d$ existing nodes to connect to. We leave this as a future work.

is the *capacity* for $(n,k,d,\alpha,\beta)$ regenerating codes (where each newcomer can access any arbitrary set of $k$ nodes).

Note that if $d<k$, requiring any $d$ storage nodes to have a flow of $\mathcal{M}$ will lead to the same condition (c.f. (14)) as requiring any $k$ storage nodes to have a flow of $\mathcal{M}$. Hence in such a case, we might as well set $k$ as $d$. For this reason, in the following we assume $d \geq k$ without loss of generality.

We are interested in characterizing the achievable tradeoffs between the storage $\alpha$ and the repair bandwidth $d\beta$. To derive the optimal tradeoffs, we can fix the repair bandwidth and solve for the minimum $\alpha$ such that (14) is satisfied. Recall that $\gamma = d\beta$ the total repair bandwidth, and the parameters $(n,k,d,\alpha,\gamma)$ can be used to characterize the system. We are interested in finding the whole region of feasible points $(\alpha,\gamma)$ and then select the one that minimizes storage $\alpha$ or repair bandwidth $\gamma$. Consider fixing both $\gamma$ and $d$ (to some integer value) and minimize $\alpha$;

$$\alpha^*(d,\gamma) \triangleq \min \quad \alpha \tag{16}$$
$$\text{subject to:} \quad \sum_{i=0}^{k-1} \min\left\{\left(1-\frac{i}{d}\right)\gamma, \alpha\right\} \geq \mathcal{M}.$$

Now observe that the dependence on $d$ must be monotone:

$$\alpha^*(d+1,\gamma) \leq \alpha^*(d,\gamma). \tag{17}$$

This is because $\alpha^*(d,\gamma)$ is always a feasible solution for the optimization for $\alpha^*(d+1,\gamma)$. Hence a larger $d$ always implies a better storage–repair bandwidth tradeoff.

The optimization (16) can be explicitly solved: We call the solution, the threshold function $\alpha^*(d,\gamma)$, which for a fixed $d$, is piecewise linear:

$$\alpha^*(d,\gamma) = \begin{cases} \frac{\mathcal{M}}{k}, & \gamma \in [f(0), +\infty) \\ \frac{\mathcal{M}-g(i)\gamma}{k-i}, & \gamma \in [f(i), f(i-1)), \end{cases} \tag{18}$$

where

$$f(i) \triangleq \frac{2\mathcal{M}d}{(2k-i-1)i + 2k(d-k+1)}, \tag{19}$$

$$g(i) \triangleq \frac{(2d-2k+i+1)i}{2d}. \tag{20}$$

The last part of the proof involves showing that the threshold function is the solution of this optimization. To simplify notation, introduce

$$b_i \triangleq \left(1-\frac{k-1-i}{d}\right)\gamma, \quad \text{for } i=0,\ldots,k-1. \tag{21}$$

Then the problem is to minimize $\alpha$ subject to the constraint:

$$\sum_{i=0}^{k-1} \min\{b_i,\alpha\} \geq B. \tag{22}$$

The left hand side of (22), as a function of $\alpha$, is a piecewise-linear function of $\alpha$:

$$\mathsf{C}(\alpha) = \begin{cases} k\alpha, & \alpha \in [0,b_0] \\ b_0 + (k-1)\alpha, & \alpha \in (b_0,b_1] \\ \vdots & \vdots \\ b_0+\ldots+b_{k-2}+\alpha, & \alpha \in (b_{k-2},b_{k-1}] \\ b_0+\ldots+b_{k-1}, & \alpha \in (b_{k-1},\infty) \end{cases}. \tag{23}$$

Note from this expression that $\mathsf{C}(\alpha)$ is strictly increasing from 0 to its maximum value $b_0+\ldots+b_{k-1}$ as $\alpha$ increases from 0 to $b_{k-1}$. To find the minimum $\alpha$ such that $\mathsf{C}(\alpha) \geq B$, we simply let $\alpha^* = \mathsf{C}^{-1}(B)$ if $B \leq b_0+\ldots+b_{k-1}$:

$$\alpha^* = \begin{cases} \frac{B}{k}, & B \in [0, kb_0] \\ \frac{B-b_0}{k-1}, & B \in (kb_0, b_0+(k-1)b_1] \\ \vdots & \vdots \\ B - \sum_{j=0}^{k-2} b_j, & B \in \left(\sum_{j=0}^{k-2} b_j + b_{k-2},\ \sum_{j=0}^{k-1} b_j\right] \end{cases} \tag{24}$$

For $i=1,\ldots,k-1$, the $i$-th condition in the above expression is:

$$\alpha* = \frac{B-\sum_{j=0}^{i-1} b_j}{k-i},$$
$$\text{for } B \in \left(\sum_{j=0}^{i-1} b_j + (k-i)b_{i-1}, \sum_{j=0}^{i} b_j + (k-i-1)b_i\right],$$

Note from the definition of $\{b_i\}$ (21) that

$$\begin{aligned}\sum_{j=0}^{i-1} b_j &= \sum_{j=0}^{i-1}\left(1-\frac{k-1-j}{d}\right)\gamma \\ &= \gamma\left[i\left(1-\frac{k-1}{d}\right)+\frac{i(i-1)}{2d}\right] \\ &= \gamma i \frac{2d-2k+i+1}{2d}, \\ &= \gamma g(i),\end{aligned}$$

and

$$\begin{aligned}&\sum_{j=0}^{i} b_j + (k-i-1)b_i \\ =&\gamma(i+1)\frac{2d-2k+i+2}{2d} + (k-i-1)\gamma\left(1-\frac{k-1-i}{d}\right) \\ =&\gamma\frac{2ik-i^2-i+2k+2kd-2k^2}{2d}, \\ =&\gamma\frac{B}{f(i)},\end{aligned}$$

where $f(i)$ and $g(i)$ are defined in (2)(3). Hence we have:

$$\alpha* = \frac{B-g(i)}{k-i}, \quad \text{for } B \in \left(\frac{\gamma B}{f(i-1)}, \frac{\gamma B}{f(i)}\right].$$

The expression of $\alpha^*(d,\gamma)$ then follows. ■